\pdfoutput=1

\documentclass[11pt]{article}

\usepackage[]{coling}

\usepackage{times}
\usepackage{latexsym}

\usepackage[T1]{fontenc}

\usepackage[utf8]{inputenc}

\usepackage{microtype}

\usepackage{inconsolata}
\usepackage{authblk}

\usepackage{graphicx}
\usepackage{algorithm}
\usepackage{algorithmic}
\usepackage{multirow}
\usepackage{subfigure}
\usepackage{amsmath}
\usepackage[most]{tcolorbox}

\title{PSM-SQL: Progressive Schema Learning with Multi-granularity Semantics for Text-to-SQL}

\author[1]{Zhuopan Yang}
\author[2]{Yuanzhen Xie}
\author[2]{Ruichao Zhong}
\author[2]{Yunzhi Tan}
\author[2]{\\Enjie Liu}
\author[1]{Zhenguo Yang}
\author[2]{Mochi Gao}
\author[2]{Bo Hu}
\author[2]{Zang Li}

\affil[1]{Guangdong University of Technology}
\affil[2]{Platform and Content Group, Tencent}

\affil[ ]{\texttt{zhuopanyang@gmail.com, \{ashexie, answerzhong, boristan, karolinaliu\}@tencent.com}}
\affil[ ]{\texttt{yzg@gdut.edu.cn, \{mochigao, harryyfhu, gavinzli\}@tencent.com}}

\begin{document}
\maketitle
\begin{abstract}
It is challenging to convert natural language (NL) questions into executable structured query language (SQL) queries for Text-to-SQL tasks due to the vast number of database schemas with redundancy, which interferes with semantic learning, and the domain shift between NL and SQL. In this paper, we propose a progressive schema linking with multi-granularity semantics (PSM-SQL) framework to reduce the redundant database schemas for Text-to-SQL. Using the multi-granularity schema linking (MSL) module, PSM-SQL learns the schema semantics at the column, table, and database levels. More specifically, a triplet loss is used at the column level to learn embeddings, while fine-tuning the LLM is employed at the database level for schema reasoning. MSL employs classifier and similarity scores to model interactions for schema linking at the table level. In particular, PSM-SQL adopts a chain loop strategy, sacrificing the upper limit of algorithm accuracy to reduce the task difficulty of schema linking by continuously decreasing the number of schemas. Experiments conducted on Text-to SQL datasets show the outperformance of the proposed PSM-SQL.
\end{abstract}

\section{Introduction} \label{sec1}
\begin{figure}[ht]
	\centering
	\includegraphics[width=0.45\textwidth]{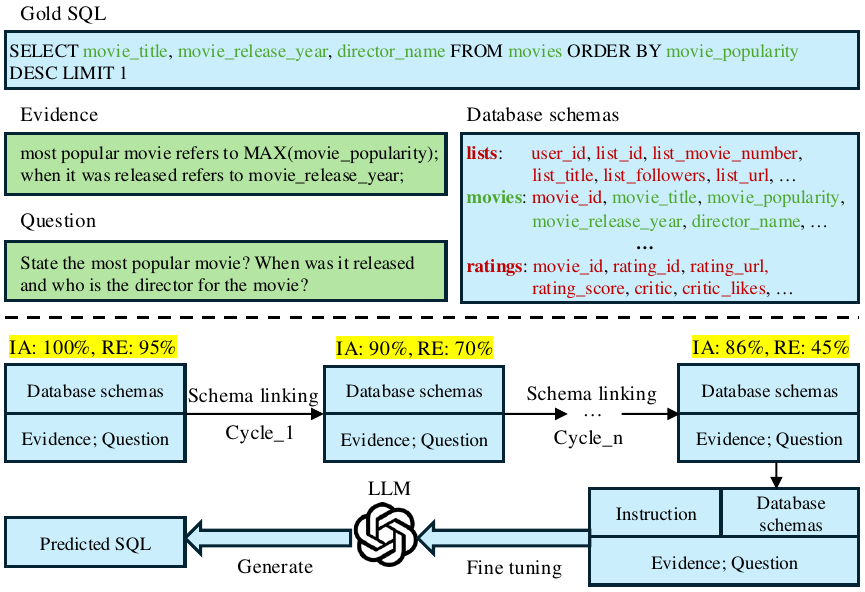}
	\caption{Illustration of an example of Text-to-SQL solved by PSM-SQL. IA and RE denote the Including Accuracy and Redundancy, respectively.} \label{img1}
\end{figure}
Text-to-SQL \cite{ref23, ref24, ref7, ref11} has attracted increasing attention in recent years, with the goal of covert natural language (NL) questions into executable structured query language (SQL) queries, thereby making it easier for users to access data in a relational database. However, the growing complexity database schemas (e.g., tables, columns, values, etc.) and domain gap between NL and SQL make it difficult to generate executable and accuracy SQL queries. By leveraging the powerful understanding and generation capabilities of large language models (LLMs), Text-to-SQL including schema linking and SQL generation tasks has made significant progress to address the problems.

The works on Text-to-SQL based on LLMs can be divided into two categories: prompting-based methods and fine-tuning-based methods. They often use domain-specific knowledge to construct prompts or fine-tune LLMs, enabling LLMs to learn the patterns for reasoning and generating SQL queries. For instance, C3 \cite{ref6}, DIN-SQL \cite{ref8}, MAC-SQL \cite{ref9} decompose the Text-to-SQL into multiple sub-problems through chains of thought (CoT) \cite{ref1}. DAIL-SQL \cite{ref7}, DTS-SQL \cite{ref11}, CODES \cite{ref12} employ supervised fine-tuning strategy to fine-tune the open-source LLMs for Text-to-SQL. However, they focus on activating and enhancing the understanding and generation capability of LLMs for Text-to-SQL task, especially for SQL generation, while neglecting the optimizations of the task itself, such as schema linking. As shown in Fig. \ref{img1}, our PSM-SQL pays attention on optimizing the task itself to continuously decrease the difficulty of Text-to-SQL. On one hand, PSM-SQL learns the semantics of schemas at the table, column, and database levels to link related schemas for generation. On the other hand, PSM-SQL iteratively performs schema linking that is chained task, with the goal of maintaining a high lookup accuracy while allowing a high degree of redundancy, thereby continuously reducing the difficulty of Text-to-SQL.

In this paper, we propose PSM-SQL for Text-to-SQL by reducing the redundant database schemas. More specifically, PSM-SQL exploit MSL module to capture the semantics of schemas at multi-granularity levels. At column level, PSM-SQL uses triplet loss to fine-tune embedding models, which rewards the correct schemas while punishes the incorrect schemas. PSM-SQL fine-tunes the LLM to reason database schemas that are related to SQL generation at database level. At the table level, PSM-SQL designs classifier and similarity scores to model schema interactions in embedding space for schema linking. In particular, a chain loop strategy is applied to continuously decrease the difficulty of tasks by reducing redundant schemas.

The contributions of this work are summarized below:
\begin{itemize}
\item[$\bullet$] We propose a progressive schema linking with multi-granularity semantics (PSM-SQL) framework for Text-to-SQL, which effectively reduce the redundant database schemas to enhance semantics learning of schemas and eliminate the domain shift between NL and SQL.
\item[$\bullet$] We devise a MSL module to learn multi-semantics of schemas at table, column, and database levels, and adopt a chain loop strategy to decrease the difficulty of schema linking continuously.
\item[$\bullet$] We conduct extensive experiments on Text-to-SQL datasets including Spider and Bird, manifesting the effectiveness of PSM-SQL against the state-of-the-art Text-to-SQL works.
\end{itemize}

The rest of the paper is organized as follows. Section \ref{sec2} reviews the related works. Section \ref{sec3} presents the details of the proposed PSM-SQL. Experiments are conducted in Section \ref{sec4}, followed by conclusions in Section \ref{sec5}.

\begin{figure*}[ht]
	\centering
	\includegraphics[width=1\textwidth]{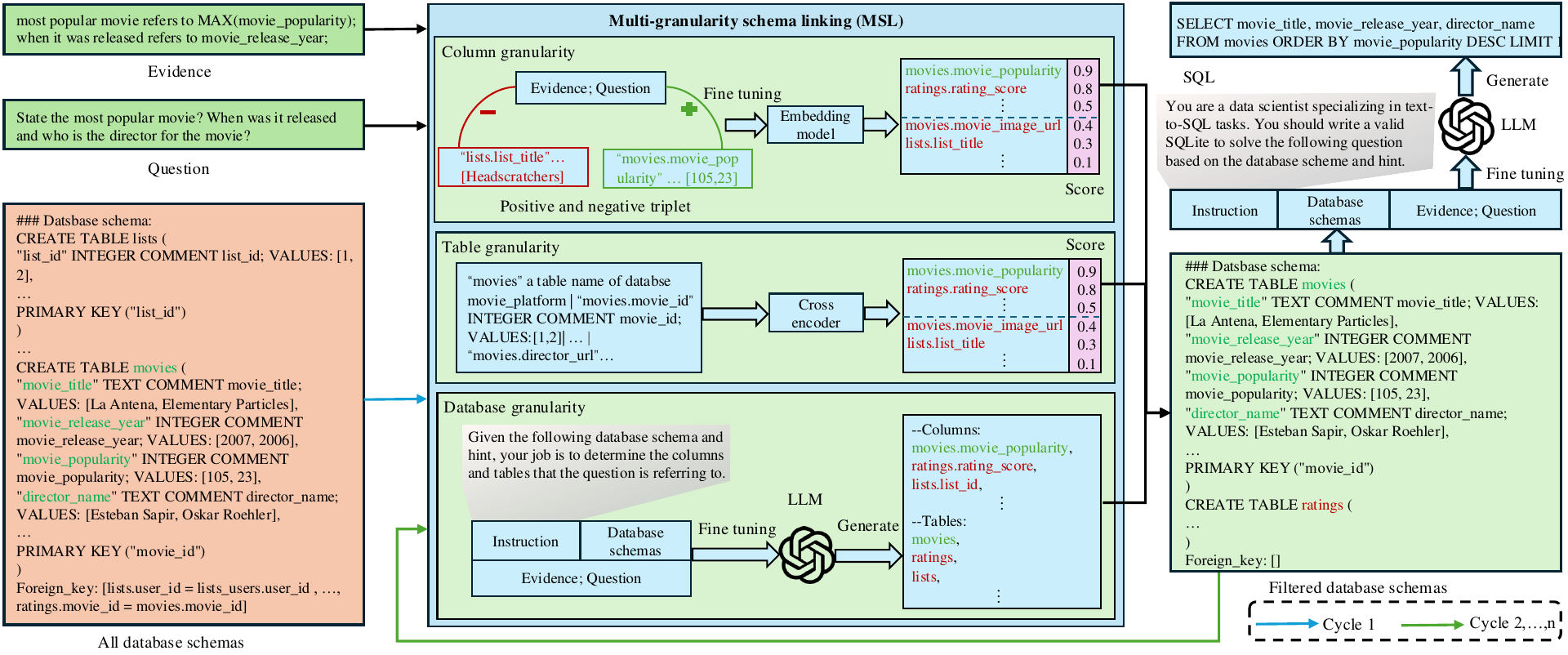}
	\caption{The framework of PSM-SQL.} \label{img2}
\end{figure*}
\section{Related Work} \label{sec2}
Text-to-SQL aims to convert natural language (NL) questions into executable structured query language (SQL) queries. In the early stage, Text-to-SQL employed encoder-decoder architectures to encode the question and database schema representations and decode the SQL queries. For example, RESDSQL \cite{ref2} proposed a ranking-enhanced encoder to select the relevant schemas and a skeleton-aware decoder to implicitly guide the SQL parsing by the skeleton. SADGA \cite{ref3} and LGESQL \cite{ref4} introduced graph neural networks to learn the relationships between questions and database schemas. Graphix-T5 \cite{ref5} designed specially-designed graph-aware layers to encode a mixture of semantic and structural information, boosting the capability of structural encoding of T5 while keeping the contextual encoding ability of the pretrained T5. However, they are suboptimal to generate SQL queries due to their limited generative capabilities caused by the small number of parameters in the model.

Recently, researcher introduced large language models (LLMs) to understand database schemas and generate SQL queries for Text-to-SQL, which can be roughly grouped into two categories: prompting-based methods and fine-tuning-based methods. More specifically, prompting-based methods designed specific prompts with contextual learning to enhance the reasoning ability of LLMs in Text-to-SQL domain. DAIL-SQL \cite{ref7} selected examples based on their skeleton similarities and removed cross domain knowledge from examples for token efficiency. DIN-SQL \cite{ref8} broke down the generation problem into sub-problems and fed the solutions of those sub-problems as prompts into LLMs to generate SQL queries. MAC-SQL \cite{ref9} employed a core decomposer agent for Text-to-SQL generation with few-shot chain-of-thought reasoning and adopt two auxiliary agents to refine erroneous SQL queries by utilizing external tools. TA-SQL \cite{ref10} proposed Task Alignment (TA) strategy to mitigate hallucinations at each stage in Text-to-SQL, reducing the burden of SQL generation. Fine-tuning-based methods used data with domain-specific knowledge to fine-tune LLMs, encouraging LLMs to learn knowledge for Text-to-SQL. DTS-SQL \cite{ref11} introduced a two-stage fine-tuning approach to decompose Text-to-SQL tasks into two simpler tasks for protecting data privacy, enabling small open-source models to rival larger ones. CODES \cite{ref12} proposed a comprehensive database prompt construction strategy and a bi-directional data augmentation method to fine-tune a series of language models ranging from 1B to 15B parameters using their collected SQL-focused corpus. However, they often focus on optimizing SQL generation while neglecting the optimization of schema linking.

\section{Problem Definition}\label{sec3}
In this section, we formalize the notations of Text-to-SQL, which aims to convert the NL questions to SQL queries. Without loss of generality, Text-to-SQL can be divided into two subtasks: schema linking and SQL generation.

Given the questions $Q=\left\{q_i\right\}_{i=1}^n$, evidence as external knowledge $K=\left\{k_i\right\}_{i=1}^n$, and the relational database $D=\left\{\left\{t_j, c_j^1, \ldots, c_j^{\left|C_j\right|}\right\}_{j=1}^{|T|}\right\}_{i=1}^n$, Text-to-SQL processes schema linking to predict the related schemas that are used for SQL generation from the original database as follows,
\begin{equation}
    D_f = Filter(Q, K, D)
\end{equation}where $Filter(\cdot)$ denotes the function of schema linking, $D_f$ denotes the filtered database that consists of the predicted schemas, $n$ denotes the number of Text-to-SQL samples, $|T|$ represents the number of tables in the relational database, $|C_j|$ is the number of columns in the $j$-th table, $t_j$ and $c_j$ is the table name and column information in the $j$-th table respectively.

Furthermore, Text-to-SQL generates the SQL queries S to answer the questions as follows,
\begin{equation}
    S = Parse(Q, K, D_f)
\end{equation}
where $Parse(\cdot)$ denotes the function of SQL generation.

\section{Methodology}\label{sec4}
\subsection{Overview of the Framework}
Given the question with evidence and database schemas, we propose a progressive schema linking with multi-granularity semantics network (denoted as PSM-SQL) as shown in Fig. \ref{img2}, which consists of a chain loop strategy and multi-granularity schema linking (MSL) module. PSM-SQL employes chain loop strategy to reducing the number of redundant schemas continuously and adopts MSL to learn the schema semantics at the column, table, and database levels. More specifically, a triplet loss is used to learn embeddings at the column level, while fine-tuned LLM is employed to reason the related schemas at the database level. At the table level, MSL adopts classifier and similarity loss to model schema interactions at the table level.

\subsection{Schema Linking}
Given the question $q \in Q$, evidence as external knowledge $k \in K$, and the relational database $d = \left\{ t_j, c_j^1, \ldots, c_j^{(|C_j|)} \right\}_{j=1}^{(|T|)} \in D$, multi-granularity schema linking (MSL) module learns the semantics of schemas at the column, table, and database levels for schema linking, capturing the multi-granularity patterns of schemas. 

\textbf{Column level.} At the column level, we construct triples $(a, c_j^p, c_j^n)$, where $a=Cat(k, q)$ to fine tune a pre-trained embedding model (e.g., BGE-large-en \cite{ref20}, etc.), ranking the scores for schemas to attain the filtered database $d_f^c$ that consist of the schemas which scores exceeding 0.5 as follows,
\begin{equation}
    d_f^c = BGE(a, c_j^p, c_j^n)
\end{equation}
where $BGE(\cdot)$ and $Cat(\cdot)$ denote the BGE operation and the concatenation operation respectively, $a$ is the anchor, $c_j^p \in d_{gt}$ and $c_j^n \notin d_{gt}$ represent the schemas from the $j$-th table that are relevant and irrelevant to generating SQL queries, and 
$d_{gt} \in D_{gt} = \left\{\left\{t_j, c_j^1, \ldots, c_j^{\left|C_j\right|}\right\}_{j=1}^{|T|}\right\}_{i=1}^n$ represents the ground-truth database schemas related to SQL generation.

The loss function at the column level can be denoted as follows,
\begin{equation}
    \mathcal L_c = \max \left( \varphi(a, c_j^p) - \varphi(a, c_j^n) + \beta, 0 \right)
\end{equation}
where $\varphi(\cdot)$ is the distance function (e.g., cosine distance, etc.), $\beta$ is the constant slack margin between the positive and negative pairs.

\begin{figure}[ht]
	\centering
	\includegraphics[width=0.48\textwidth]{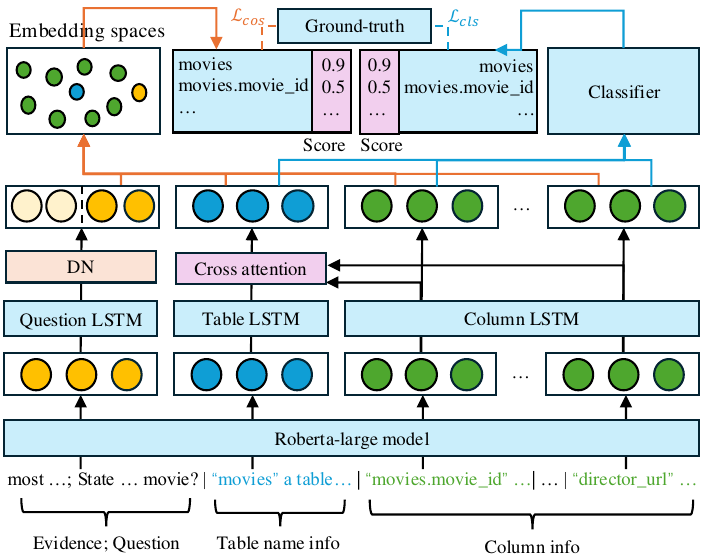}
	\caption{Details of the Multi-granularity schema linking (MSL) module.} \label{img3}
\end{figure}

\textbf{Table level.} As shown in Fig. \ref{img3}, a cross encoder is implemented to model the interaction of schemas in the embedding and classifier spaces. Given the question with evidence $a=Cat(k, q)$, and the table schemas $\left\{ t_j, c_j^1, \ldots, c_j^{(|C_j|)} \right\} \in d$, we employ the pre-trained ROBERTA model \cite{ref21} to obtain the sequence embeddings, and then attain fusion embeddings of question, table and columns by long short-term memory (LSTM) network respectively as follows,
\begin{equation}
    e_q, e_t, \left\{ e_c^k \right\}_{k=1}^{|C_j|} = EMB(a, t_j, c_j^1, \ldots, c_j^{(|C_j|)})
\end{equation}
where $EMB(\cdot)=LSTM(ROBERTA(\cdot))$ is the function that encodes the feature embeddings, $ROBERTA(\cdot)$ and $LSTM(\cdot)$ is the ROBEARTA and LSTM operation, $e_q$ and $e_t$ denote the question and table embeddings, $\left\{ e_c^k \right\}_{k=1}^{|C_j|}$ represent the column embeddings.

In particular, we adopt a disentangled network (DN) that is a fully connected layer with dropout operation to filter the unrelated semantics of question embeddings from redundant tokens (e.g., “the”, “was”, etc.) as follows,
\begin{equation}
    \left[ e_q^n; e_q^s \right] = \text{ReLU}\left( \text{DN}\left( e_q \right) \right)
\end{equation}
where $ReLU(\cdot)$ is the ReLU activation function, $e_q^n$ and $e_q^s$ denote the question embeddings that are semantically unrelated and semantically related to SQL generation respectively.

To attain column-enhanced table embeddings $e_t^c$, a multi-head scaled dot-product attention layer, and a feature fusion layer are exploited to fuse the semantics of columns into table embeddings as follows,
\begin{equation}
    e_t^a = \text{MultiHeadAttn}\left( e_t, \left\{ e_c^k \right\}_{k=1}^{\left| C_j \right|}, \left\{ e_c^k \right\}_{k=1}^{\left| C_j \right|}, h \right)
\end{equation}
\begin{equation}
    e_t^c = \text{Norm}(e_t, e_t^a)
\end{equation}
where $MultiHeadAttn(\cdot)$ represents the multi-head attention function, h is number of heads, and $Norm(\cdot)$ is a row-wise L\_2   normalization function.

Furthermore, we use distance function $\varphi(\cdot)$ (e.g., cosine distance, etc.) and classifier $Classifier(\cdot)$ to obtain the scores of schemas which represents the probability belong to ground-truth schemas from model as follows,
\begin{equation}
    \text{score}_{\text{cos}} = 1 - \varphi\left( e_q^s, \left\{ e_t^c, \left\{ e_c^k \right\}_{k=1}^{\left| C_j \right|} \right\} \right)
\end{equation}
\begin{equation}
    \text{score}_{\text{cl}} = \text{Classifier}\left( e_q^s, \left\{ e_t^c, \left\{ e_c^k \right\}_{k=1}^{\left| C_j \right|} \right\} \right)
\end{equation}
where $\text{score}_{\text{cos}}$ and $\text{score}_{\text{cl}}$ are the scores of schemas obtained by cosine similarity distance and classifier respectively.

Finally, we retain database schemas with scores exceeding 0.5 as the predicted database schemas $d_{pred}^{cos}$ and $d_{pred}^{cl}$ from $\text{score}_{\text{cos}}$ and $\text{score}_{\text{cl}}$ respectively. For the inference, the predicted schemas of cross encoder are $d_f^t = d_{\text{pred}}^{\text{cos}} \cup d_{\text{pred}}^{\text{cl}}$. Thus, we use cross entropy loss $CrossEntropy(\cdot)$ to train the cross encoder from the perspectives of cosine similarity distance and classifier as follows,
\begin{equation}
    \mathcal L_{\text{cos}} = \text{CrossEntropy}\left( d_{\text{pred}}^{\text{cos}}, d_{\text{gt}} \right)
\end{equation}
\begin{equation}
    \mathcal L_{\text{cl}} = \text{CrossEntropy}\left( d_{\text{pred}}^{\text{cl}}, d_{\text{gt}} \right)
\end{equation}
where $d_{gt} \in D_{gt}$ is the ground-truth database schemas.

The loss function at the table level can be denoted as follows,
\begin{equation}
    \mathcal L_t = \mathcal L_{cos} + \mathcal L_{cl}
\end{equation}

\textbf{Database level.} 
Given the question $q \in Q$, evidence as external knowledge $k \in K$, and the relational database $d \in D$, we construct an instruction to fine tune pre-trained LLMs (e.g., Llama3-8B, etc.) using Lora \cite{ref22}, reasoning the predicted schemas related to SQL queries for schema linking at the database level as follows,
\begin{equation}
    d_f^d = \text{LLM}(inst, d, k, q)
\end{equation}
where $inst$ is the instruction, $LLM(\cdot)$ denotes the LLM operation, and $d_f^d$ represents the filtered database schemas at the database level.

The loss function at the database level can be denoted as follows,
\begin{equation}
    \mathcal L_d = \frac{1}{n} \sum_{i=1}^{n} p\left( d_{\text{gt}}^i \middle| d_i, k_i, q_i \right)
\end{equation}
where $n$ is the number of Text-to-SQL instances, $d_i$, $k_i$, $q_i$ are the relational database, the evidence and the question of the $i$-th instance, $d_{gt}^i$ denotes the $i$-th target database schemas related to SQL generation.

Finally, we attain the filtered database schemas $d_f = d_f^c \cup d_f^t \cup d_f^d$, which serves as the input database for schema linking of the next cycle. In particular, we follow the chain loop strategy to process schema linking using MSL continuously, reducing the task difficulty of schema linking by continuously decreasing the number of redundant schemas.

\subsection{SQL Generation}
Given the predicted database schemas $d_f = d_f^c \cup d_f^t \cup d_f^d$ from MSL, we construct an instruction to fine tune pre-trained LLMs using Lora to generate SQL query for the inference as follows,
\begin{equation}
    s = \text{LLM}(inst, d_f, k, q)
\end{equation}
where $s$ is the predicted SQL query.

The loss function of SQL generation can be denoted as follows,
\begin{equation}
    \mathcal L_SQL = \frac{1}{n} \sum_{i=1}^{n} p\left( s_{\text{gt}}^i \middle| d_f^i, k_i, q_i \right)
\end{equation}
where $n$ is the number of Text-to-SQL instances, $s_{gt}^i$ is the $i$-th target ground-truth SQL query.

\section{Experiments}\label{sec5}
\subsection{Datasets}
We conduct evaluations on two widely used Text-to-SQL datasets including Spider \cite{ref13} and Bird \cite{ref14}.

\textbf{1) Spider \cite{ref13}.} It contains 200 database schemas, of which 160 database chemas are used for training and validation, and 40 database schemas are used for testing. The training set comprises 8,659 samples, including 7,000 manually annotated samples and 1,659 samples sourced from six previous Text-to-SQL datasets (e.g., Restaurants \cite{ref15, ref16}, GeoQuery \cite{ref17}, etc.) for supplement, while the development set contains 1,034 samples, and the test set is hidden. Each instance is consisted of a NL question on specific database schemas and a corresponding SQL query.

\textbf{2) Bird \cite{ref14}.} It covers 95 large databases consisting of 69, 11, and 15 databases for training, development, and testing respectively. This dataset with a combined size of 33.4 GB covers 37 professional domains, including blockchain, sports, health care, etc. The training set has 9,428 samples, while the development set includes 1,543 samples, and the test set is hidden. Each sample consists of a NL question with an evidence as external knowledge, a specific database and a corresponding SQL query.

\subsection{Evaluation Metrics}
To make a complete comparison, we evaluate the performances on schema linking and SQL generation for Text-to-SQL.

\textbf{1) Schema Linking.} In terms of evaluations on schema linking, we adopt matching accuracy (MA), Including accuracy (IA), and redundancy (RE) to report the performances.

\begin{table}[]
\caption{Compared with the state-of-the-art Text-to-SQL methods on Bird Dataset. * indicates the results obtained by ourselves with the codes released by the authors. The best results are marked in bold.}
\label{table1}
\scriptsize
\renewcommand\arraystretch{1.3}
\setlength{\tabcolsep}{2.4pt}
\begin{tabular}{|l|l|cc|cc|}
\hline
                                                                                     & Method                                                                                  & \multicolumn{2}{c|}{Dev}                             & \multicolumn{2}{c|}{Test}          \\ \hline
                                                                                     &                                                                                         & \multicolumn{1}{c|}{EX}             & VES             & \multicolumn{1}{c|}{EX}    & VES    \\ \hline
\multirow{10}{*}{\begin{tabular}[c]{@{}l@{}}Prompting \\ -based \\ methods\end{tabular}} & GPT-4                                                                                   & \multicolumn{1}{c|}{46.35}          & 49.77          & \multicolumn{1}{c|}{54.89} & 60.77 \\ \cline{2-6} 
                                                                                     & \begin{tabular}[c]{@{}l@{}}DIN-SQL+GPT-4\\ \cite{ref8}\end{tabular}                                                             & \multicolumn{1}{c|}{50.72}          & 58.79          & \multicolumn{1}{c|}{55.90} & 59.44 \\ \cline{2-6} 
                                                                                     & \begin{tabular}[c]{@{}l@{}}DAIL-SQL+GPT-4 \\ \cite{ref7}\end{tabular}                                                                         & \multicolumn{1}{c|}{54.76}          & 56.08          & \multicolumn{1}{c|}{57.41} & 61.95 \\ \cline{2-6} 
                                                                                     & \begin{tabular}[c]{@{}l@{}}MCS-SQL+GPT-4 \\ \cite{ref18}\end{tabular}                                                                           & \multicolumn{1}{c|}{63.4}           & 64.8           & \multicolumn{1}{c|}{65.5}  & 71.4  \\ \cline{2-6} 
                                                                                     & \begin{tabular}[c]{@{}l@{}}MAC-SQL+GPT-4 \\ \cite{ref9}\end{tabular}                                                                           & \multicolumn{1}{c|}{59.39}          & 66.39          & \multicolumn{1}{c|}{59.59} & 67.68 \\ \cline{2-6} 
                                                                                     & \begin{tabular}[c]{@{}l@{}}MAC-SQL+GPT-3.5-Trubo \\ \cite{ref9}\end{tabular}                                                                   & \multicolumn{1}{c|}{50.56}          & 61.25          & \multicolumn{1}{c|}{-}     & -     \\ \cline{2-6} 
                                                                                     & MAC-SQL+GPT-3.5-Trubo*                                                                  & \multicolumn{1}{c|}{49.15}          & 56.93          & \multicolumn{1}{c|}{-}     & -     \\ \cline{2-6} 
                                                                                     & \begin{tabular}[c]{@{}l@{}}MAC-SQL+GPT-3.5-Trubo*\\ +Our predicted schemas\end{tabular} & \multicolumn{1}{c|}{52.28}          &  56.14              & \multicolumn{1}{c|}{-}     & -     \\ \cline{2-6} 
                                                                                     & MAC-SQL+Llama3-70B*                                                                      & \multicolumn{1}{c|}{53.72}          & 57.03          & \multicolumn{1}{c|}{-}     & -     \\ \cline{2-6} 
                                                                                     & \begin{tabular}[c]{@{}l@{}}MAC-SQL+Llama3-70B*\\ +Our predicted schemas\end{tabular}     & \multicolumn{1}{c|}{54.95}          & 58.76          & \multicolumn{1}{c|}{-}     & -     \\ \hline
\multirow{7}{*}{\begin{tabular}[c]{@{}l@{}}Fine-tuning \\ -based\\ methods\end{tabular}} & \begin{tabular}[c]{@{}l@{}}DTS-SQL+DeepSeek-7B \\ \cite{ref11}\end{tabular}                                                                     & \multicolumn{1}{c|}{55.8}           & -              & \multicolumn{1}{c|}{60.31} & -     \\ \cline{2-6} 
                                                                                     & SFT CODES-7B \cite{ref12}                                                                            & \multicolumn{1}{c|}{57.17}          & 58.80          & \multicolumn{1}{c|}{59.25} & 63.62 \\ \cline{2-6} 
                                                                                     & \begin{tabular}[c]{@{}l@{}}SFT CODES-15B \\ \cite{ref12}\end{tabular}                                                                           & \multicolumn{1}{c|}{58.47}          & 59.87          & \multicolumn{1}{c|}{60.37} & 64.22 \\ \cline{2-6} 
                                                                                     & CHESS \cite{ref19}                                                                                  & \multicolumn{1}{c|}{65.00}          & 65.43          & \multicolumn{1}{c|}{66.69} & 72.63 \\ \cline{2-6} 
                                                                                     & PSM-SQL+DeepSeek-7B (Ours)                                                              & \multicolumn{1}{c|}{60.30}          & 61.57          & \multicolumn{1}{c|}{-}     & -     \\ \cline{2-6} 
                                                                                     & PSM-SQL+Llama3-8B (Ours)                                                                & \multicolumn{1}{c|}{59.71}          & 65.15          & \multicolumn{1}{c|}{-}     & -     \\ \cline{2-6} 
                                                                                     & PSM-SQL+Llama3-70B (Ours)                                                               & \multicolumn{1}{c|}{\textbf{65.06}} & \textbf{67.87} & \multicolumn{1}{c|}{-}     & -     \\ \hline
\end{tabular}
\end{table}

\textbf{Matching Accuracy:} The MA metric is used to evaluate whether the predicted schemas are the same as the ground-truth schemas.
\begin{equation}
M A=\frac{\sum_n^i 1 \text { if } \operatorname{match}\left(s_i, y_i\right)=\text { True else } 0}{n}
\end{equation}
where n denotes the number of samples, $s_i$ and $y_i$ represent the $i$-th predicted and ground-truth schemas respectively, match($\cdot$) denotes the function that is used to determine if the two schemas are the same.
	
\textbf{Including Accuracy:} The IA metric is employed to evaluate whether the predicted schemas contain the ground-truth schemas.
\begin{equation}
I A=\frac{\sum_n^i 1 \text { if } s_i \supseteq y_i \text { else } 0}{n}
\end{equation}

\textbf{Redundancy:} The RE metric evaluates the redundancy in the predicted schemas.
\begin{equation}
R E=\frac{\sum_n^i \sum_m^j 0 \text { if } s_i^j \in y_i \text { else } 1}{n \times m}
\end{equation}
where m denotes the number of schemas in the $i$-th predicted schemas and $s_i^j$ represents the $j$-th schema in the $i$-th predicted schemas.

\textbf{2) SQL Generation.} For Spider dataset, we use execution accuracy (EX) and exact match accuracy (EM) to evaluate the performances. In terms of the evaluations on Bird dataset, we employ execution accuracy (EX) and valid efficiency score (VES) as the evaluation metrics. 

\begin{table}[]
\caption{Compared with the state-of-the-art Text-to-SQL methods on Spider Dataset. * indicates the results obtained by ourselves with the codes released by the authors. The best results are marked in bold.}
\scriptsize
\renewcommand\arraystretch{1.3}
\setlength{\tabcolsep}{2.7pt}
\label{table2}
\begin{tabular}{|l|l|cc|cc|}
\hline
                                                                                     & Method                                                                                  & \multicolumn{2}{c|}{Dev}           & \multicolumn{2}{c|}{Test}         \\ \hline
                                                                                     &                                                                                         & \multicolumn{1}{c|}{EX}    & EM    & \multicolumn{1}{c|}{EX}    & EM   \\ \hline
\multirow{10}{*}{\begin{tabular}[c]{@{}l@{}}Prompting \\ -based \\ methods\end{tabular}} & C3+ChatGPT \cite{ref6}                                                                             & \multicolumn{1}{c|}{81.80} & -     & \multicolumn{1}{c|}{82.30} & -    \\ \cline{2-6} 
                                                                                     & \begin{tabular}[c]{@{}l@{}}DIN-SQL+GPT-4 \\ \cite{ref8}\end{tabular}                                                                          & \multicolumn{1}{c|}{74.2}  & 60.1  & \multicolumn{1}{c|}{85.30} & 60   \\ \cline{2-6} 
                                                                                     & \begin{tabular}[c]{@{}l@{}}DAIL-SQL+GPT-4 \\ \cite{ref7}\end{tabular}                                                                         & \multicolumn{1}{c|}{84.40} & 74.4  & \multicolumn{1}{c|}{86.60} & -    \\ \cline{2-6} 
                                                                                     & \begin{tabular}[c]{@{}l@{}}MCS-SQL+GPT-4 \\ \cite{ref18}\end{tabular}                                                                          & \multicolumn{1}{c|}{\textbf{89.5}}  & -     & \multicolumn{1}{c|}{89.6}  & -    \\ \cline{2-6} 
                                                                                     & \begin{tabular}[c]{@{}l@{}}MAC-SQL+GPT-4 \\ \cite{ref9}\end{tabular}                                                                          & \multicolumn{1}{c|}{86.75} & 63.20 & \multicolumn{1}{c|}{82.80} & -    \\ \cline{2-6} 
                                                                                     & \begin{tabular}[c]{@{}l@{}}MAC-SQL+GPT-3.5-Trubo \\ \cite{ref9}\end{tabular}                                                                  & \multicolumn{1}{c|}{80.56} & -     & \multicolumn{1}{c|}{75.53} & -    \\ \cline{2-6} 
                                                                                     & MAC-SQL+GPT-3.5-Trubo*                                                                  & \multicolumn{1}{c|}{75.0}  & 19.1  & \multicolumn{1}{c|}{-}     & -    \\ \cline{2-6} 
                                                                                     & \begin{tabular}[c]{@{}l@{}}MAC-SQL+GPT-3.5-Trubo*\\ +Our predicted schemas\end{tabular} & \multicolumn{1}{c|}{75.4}  & 20.5  & \multicolumn{1}{c|}{-}     & -    \\ \cline{2-6} 
                                                                                     & MAC-SQL+Llama3-70B*                                                                      & \multicolumn{1}{c|}{76.6}  & 27.5  & \multicolumn{1}{c|}{-}     & -    \\ \cline{2-6} 
                                                                                     & \begin{tabular}[c]{@{}l@{}}MAC-SQL+Llama3-70B*\\ +Our predicted schemas\end{tabular}     & \multicolumn{1}{c|}{78.1}  & 28.7  & \multicolumn{1}{c|}{-}     & -    \\ \hline
\multirow{8}{*}{\begin{tabular}[c]{@{}l@{}}Fine-tuning \\ -based\\ methods\end{tabular}} & RESDSQL-3B \cite{ref2}                                                                             & \multicolumn{1}{c|}{84.1}  & 80.5  & \multicolumn{1}{c|}{79.9}  & 72   \\ \cline{2-6} 
                                                                                     & \begin{tabular}[c]{@{}l@{}}DTS-SQL+DeepSeek-7B \\ \cite{ref11}\end{tabular}                                                                    & \multicolumn{1}{c|}{85.5}  & 79.1  & \multicolumn{1}{c|}{84.4}  & 73.7 \\ \cline{2-6} 
                                                                                     & SFT CODES-7B \cite{ref12}                                                                           & \multicolumn{1}{c|}{85.5}  & -     & \multicolumn{1}{c|}{-}     & -    \\ \cline{2-6} 
                                                                                     & SFT CODES-15B \cite{ref12}                                                                          & \multicolumn{1}{c|}{84.9}  & -     & \multicolumn{1}{c|}{-}     & -    \\ \cline{2-6} 
                                                                                     & CHESS \cite{ref19}                                                                                  & \multicolumn{1}{c|}-  & -     & \multicolumn{1}{c|}{87.2}     & -    \\ \cline{2-6} 
                                                                                     & PSM-SQL+DeepSeek-7B (Ours)                                                              & \multicolumn{1}{c|}{86.9}  & 79.5  & \multicolumn{1}{c|}{-}      &-      \\ \cline{2-6} 
                                                                                     & PSM-SQL+Llama3-8B (Ours)                                                                & \multicolumn{1}{c|}{84.1}      &79.5       & \multicolumn{1}{c|}{-}      &-      \\ \cline{2-6} 
                                                                                     & PSM-SQL+Llama3-70B (Ours)                                                               & \multicolumn{1}{c|}{88.6}      &\textbf{82.2}       & \multicolumn{1}{c|}{-}      &-      \\ \hline
\end{tabular}
\end{table}

\textbf{Execution Accuracy:} The EX metric evaluates whether the predicted and ground-truth SQL queries yield the same execution results on the database.
\begin{scriptsize}
\begin{equation}
E X=\frac{\sum_n^i 1 \text { if match}\left(\text {ex}\left(S Q L_i^{\text {pred}}\right), \text {ex}\left(S Q L_i^{g t}\right)\right)=\text {True else } 0}{n}
\end{equation}
\end{scriptsize}where ex ($\cdot$) is the function that is used to execute the SQL queries in the database and return the results, $(S Q L_i^{\text {pred }}$ and $(S Q L_i^{\text {gt }}$ denote the $i$-th predicted and ground-truth SQL queries respectively.

\textbf{Valid Efficiency Score:} The VES metric evaluates the execution efficiency of accurately generated SQL queries.
\begin{equation}
V E S=\frac{E X \times \sum_n^i \frac{\operatorname{time}\left(e x\left(S Q L_i^{g t}\right)\right)}{\operatorname{time}\left(e x\left(S Q L_i^{p r e d}\right)\right)}}{n}
\end{equation}
where time ($\cdot$) denotes the execution time.

\begin{table*}[]
\caption{Compared with the state-of-the-art Text-to-SQL methods on schema linking task. * indicates the results obtained by ourselves with the codes released by the authors. The best results are in bold.}
\centering
\scriptsize
\renewcommand\arraystretch{1.3}
\setlength{\tabcolsep}{3pt}
\label{table3}
\begin{tabular}{|l|cccccc|cccccc|}
\hline
                        & \multicolumn{6}{c|}{The dev set of Bird}                                                                                                                                                                   & \multicolumn{6}{c|}{The dev set of Spider}                                                                                                                                                                \\ \hline
\multirow{2}{*}{Method} & \multicolumn{3}{c|}{Table}                                                                                    & \multicolumn{3}{c|}{Column}                                                                & \multicolumn{3}{c|}{Table}                                                                                    & \multicolumn{3}{c|}{Column}                                                               \\ \cline{2-13} 
                        & \multicolumn{1}{c|}{MA$\uparrow$}             & \multicolumn{1}{c|}{IA$\uparrow$}           & \multicolumn{1}{c|}{RE$\downarrow$}             & \multicolumn{1}{c|}{MA$\uparrow$}             & \multicolumn{1}{c|}{IA$\uparrow$}             & \multicolumn{1}{c|}{RE$\downarrow$}             & \multicolumn{1}{c|}{MA$\uparrow$}             & \multicolumn{1}{c|}{IA$\uparrow$}           & \multicolumn{1}{c|}{RE$\downarrow$}             & \multicolumn{1}{c|}{MA$\uparrow$}             & \multicolumn{1}{c|}{IA$\uparrow$}             & \multicolumn{1}{c|}{RE$\downarrow$}            \\ \hline
MAC-SQL+GPT3.5-Trubo*   & \multicolumn{1}{c|}{1.04}           & \multicolumn{1}{c|}{\textbf{100}} & \multicolumn{1}{c|}{72.77}          & \multicolumn{1}{c|}{0}              & \multicolumn{1}{c|}{81.68}          & 88.21          & \multicolumn{1}{c|}{5.22}           & \multicolumn{1}{c|}{\textbf{100}} & \multicolumn{1}{c|}{66.54}          & \multicolumn{1}{c|}{0}              & \multicolumn{1}{c|}{88.88}          & 93.36         \\ \hline
MAC-SQL+Llama3-70B*     & \multicolumn{1}{c|}{1.04}           & \multicolumn{1}{c|}{\textbf{100}} & \multicolumn{1}{c|}{72.77}          & \multicolumn{1}{c|}{0}              & \multicolumn{1}{c|}{\textbf{89.31}} & 89.97          & \multicolumn{1}{c|}{5.22}           & \multicolumn{1}{c|}{\textbf{100}} & \multicolumn{1}{c|}{66.54}          & \multicolumn{1}{c|}{0}              & \multicolumn{1}{c|}{91.88}          & 93.33         \\ \hline
PSM-SQL+Llama3-8B (Ours)                 & \multicolumn{1}{c|}{\textbf{56.65}} & \multicolumn{1}{c|}{97.91}        & \multicolumn{1}{c|}{\textbf{25.43}} & \multicolumn{1}{c|}{\textbf{17.99}} & \multicolumn{1}{c|}{86.05}          & \textbf{45.66} & \multicolumn{1}{c|}{\textbf{83.08}} & \multicolumn{1}{c|}{98.84}        & \multicolumn{1}{c|}{\textbf{10.54}} & \multicolumn{1}{c|}{\textbf{59.96}} & \multicolumn{1}{c|}{\textbf{95.16}} & \textbf{23.1} \\ \hline
\end{tabular}
\end{table*}

\textbf{Exact Match Accuracy:} The EM metric evaluates whether the abstract syntax tree structure of predicted and ground-truth SQL are the same.
\begin{scriptsize}
\begin{equation}
E X=\frac{\sum_n^i 1 \text { if match}\left(\text {st}\left(S Q L_i^{\text {pred}}\right), \text {st}\left(S Q L_i^{g t}\right)\right)=\text {True else } 0}{n}
\end{equation}
\end{scriptsize}where st ($\cdot$) is the function that is used to attain the abstract syntax tree structure of SQL queries.

\begin{table}[]
\caption{Performance of variations on the Bird Dataset. The best results are marked in bold.}
\scriptsize
\renewcommand\arraystretch{1.3}
\setlength{\tabcolsep}{3.3pt}
\label{table4}
\begin{tabular}{|l|l|cccccc|}
\hline
\multirow{3}{*}{Cycle}    & \multirow{3}{*}{}                                            & \multicolumn{6}{c|}{The dev set of Bird}                                                                                                                                                                   \\ \cline{3-8} 
                          &                                                              & \multicolumn{3}{c|}{Table}                                                                                    & \multicolumn{3}{c|}{Column}                                                                \\ \cline{3-8} 
                          &                                                              & \multicolumn{1}{c|}{MA$\uparrow$}           & \multicolumn{1}{c|}{IA$\uparrow$}            & \multicolumn{1}{c|}{RE$\downarrow$}           & \multicolumn{1}{c|}{MA$\uparrow$}            & \multicolumn{1}{c|}{IA$\uparrow$}            & \multicolumn{1}{c|}{RE$\downarrow$}            \\ \hline
\multirow{5}{*}{Cycle\_1} & Cross\_encoder                                               & \multicolumn{1}{c|}{37.87}         & \multicolumn{1}{c|}{82.4}           & \multicolumn{1}{c|}{36.29}         & \multicolumn{1}{c|}{15.45}          & \multicolumn{1}{c|}{48.63}          & 48.73          \\ \cline{2-8} 
                          & Emb\_LLM                                                     & \multicolumn{1}{c|}{28.81}         & \multicolumn{1}{c|}{89.96}          & \multicolumn{1}{c|}{48.18}         & \multicolumn{1}{c|}{5.93}           & \multicolumn{1}{c|}{46.41}          & 68.53          \\ \cline{2-8} 
                          & Gen\_LLM                                                     & \multicolumn{1}{c|}{68.12}         & \multicolumn{1}{c|}{67.34}          & \multicolumn{1}{c|}{6.68}          & \multicolumn{1}{c|}{30.7}           & \multicolumn{1}{c|}{33.05}          & 33.82          \\ \cline{2-8} 
                          & \begin{tabular}[c]{@{}l@{}}Cross\_encoder\\ +Gen\_LLM\end{tabular} & \multicolumn{1}{c|}{39.31}         & \multicolumn{1}{c|}{91.07}          & \multicolumn{1}{c|}{35.49}         & \multicolumn{1}{c|}{13.17}          & \multicolumn{1}{c|}{66.36}          & 50.44          \\ \cline{2-8} 
                          & MSL (All)                                                          & \multicolumn{1}{c|}{24.32}         & \multicolumn{1}{c|}{96.81}          & \multicolumn{1}{c|}{50.87}         & \multicolumn{1}{c|}{2.09}           & \multicolumn{1}{c|}{74.9}           & 69.07          \\ \hline
\multirow{5}{*}{Cycle\_2} & Cross\_encoder                                               & \multicolumn{1}{c|}{49.54}         & \multicolumn{1}{c|}{81.1}           & \multicolumn{1}{c|}{27.43}         & \multicolumn{1}{c|}{17.8}           & \multicolumn{1}{c|}{55.02}          & 48.99          \\ \cline{2-8} 
                          & Emb\_LLM                                                     & \multicolumn{1}{c|}{34.16}         & \multicolumn{1}{c|}{79.86}          & \multicolumn{1}{c|}{41.64}         & \multicolumn{1}{c|}{8.15}           & \multicolumn{1}{c|}{36.9}           & 62.96          \\ \cline{2-8} 
                          & Gen\_LLM                                                     & \multicolumn{1}{c|}{\textbf{94.2}} & \multicolumn{1}{c|}{94.65}          & \multicolumn{1}{c|}{\textbf{2.11}} & \multicolumn{1}{c|}{\textbf{54.95}} & \multicolumn{1}{c|}{62.97}          & \textbf{11.16} \\ \cline{2-8} 
                          & \begin{tabular}[c]{@{}l@{}}Cross\_encoder\\ +Gen\_LLM\end{tabular} & \multicolumn{1}{c|}{56.65}         & \multicolumn{1}{c|}{97.91}          & \multicolumn{1}{c|}{25.43}         & \multicolumn{1}{c|}{17.99}          & \multicolumn{1}{c|}{86.05}          & 45.66          \\ \cline{2-8} 
                          & MSL (All)                                                          & \multicolumn{1}{c|}{32.79}         & \multicolumn{1}{c|}{\textbf{98.63}} & \multicolumn{1}{c|}{43.07}         & \multicolumn{1}{c|}{3.26}           & \multicolumn{1}{c|}{\textbf{89.05}} & 62.32          \\ \hline
\end{tabular}
\end{table}

\begin{table}[]
\caption{Performance of variations on the Bird Dataset. The best results are in bold.}
\centering
\scriptsize
\renewcommand\arraystretch{1.3}
\setlength{\tabcolsep}{3.5pt}
\label{table5}
\begin{tabular}{|l|l|cc|}
\hline
\multirow{2}{*}{Cycle}    & \multirow{2}{*}{} & \multicolumn{2}{c|}{The dev set of Bird}             \\ \cline{3-4} 
                          &                   & \multicolumn{1}{c|}{EX}             & VES            \\ \hline
\multirow{2}{*}{Cycle\_1} & Cross\_encoder+Gen\_LLM & \multicolumn{1}{c|}{57.11}          & 58.7           \\ \cline{2-4} 
                          & MSL (All)               & \multicolumn{1}{c|}{57.63}          & 59.76          \\ \hline
\multirow{2}{*}{Cycle\_2} & Cross\_encoder+Gen\_LLM & \multicolumn{1}{c|}{\textbf{59.71}} & \textbf{65.15} \\ \cline{2-4} 
                          & MSL (All)               & \multicolumn{1}{c|}{58.80}          & 53.28          \\ \hline
\end{tabular}
\end{table}

\subsection{Baselines}
We include a number of 9 Text-to-SQL approaches as baselines in two categories including prompting-based methods and fine-tuning-based methods. The details are as follows.

\textbf{1) Prompting-based methods:} C3 \cite{ref6} proposed three key components including clear prompting, calibration with hints, and consistent output to improve the Text-to-SQL performance from the perspectives of model input, model bias, and model output. DIN-SQL \cite{ref8} decomposed the task of Text-to-SQL into multiple sub-tasks to address the disparity between NL and SQL. DAIL-SQL \cite{ref7} proposed a novel in-context learning technique to address the trade-off between example quality and quantity by learning the mapping between questions and queries. MCS-SQL \cite{ref18} leveraged multiple prompts to generate various candidate answers and effectively aggregated them based on confidence scores. MAC-SQL \cite{ref9} exploited LLMs as intelligent agents with different functionalities for Text-to-SQL parsing.

\textbf{2) Fine-tuning-based methods:} RESDSQL \cite{ref2} employed a skeleton-aware decoder to implicitly guide the SQL parsing by the skeleton and relevant schemas, which are selected by a ranking-enhanced encoder. DTS-SQL \cite{ref11} broke down the Text-to-SQL tasks into two simple tasks and fine-tuned the open-source LLMs to generate SQL queries. Using the collected SQL-focused corpus, CODES \cite{ref12} presented a database prompt construction strategy and a bi-directional data augmentation mechanism to fine-tune a series of languages models ranging from 1B to 15B parameters. CHESS \cite{ref19} introduced a hierarchical retrieval method to select relevant entities and contexts, and a three-staged schema pruning protocol to extract a minimally sufficient schema.

\subsection{Implementation Details}
In terms of the column level of MSL module, a pre-trained BGE-large-en-v1.5 model is adopted as the embedding model for fine-tuning, and we take 0.5 as the threshold to predict the schemas during inference stage. Regarding the table level of MSL module, we design a cross-encoder model based on ROBERTA to model the interactions between schemas in the embedding and classifier spaces. Specifically, the part of the input tokens that exceeds 512 are truncated due to the ROBERTA, which interferes with the semantic learning. To address this, the order of columns in a table will be shuffled in the input batch during training stage. During inference stage, cross-encoder first selects tables with scores greater than 0.5. Then we select the Top-8 columns for the Top-2 tables, and for the remaining tables, we select the Top-4 columns, as the predicted schemas. In terms of the database level of MSL module, we use pre-trained Llama3-8B model to fine-tune for reasoning the schemas related to SQL generation.

\subsection{Performance of the Tested Approaches}
Table \ref{table1} and Table \ref{table2} summarize the performance of the approaches on the Bird and Spider datasets in terms of SQL generation, from which we have some observations. 1) As prompting-based methods, MAC-SQL with our predicted schemas achieves better performance than the original MC-SQL on the EX and VES metrics. The reason is that the predicted schemas by ours contain less redundant schemas and a smaller input token, causing less interference with the SQL generation of LLMs. 2) As fine-tune-based methods using DeepSeek-7B, PSM-SQL+DeepSeek-7B achieves better performance over DTS-SQL+DeepSeek-7B on the EX and VES metrics. The reason is that the predicted schemas with less redundant information can benefit LLMs to learn to generate the correct SQL queries. 3) PSM-SQL+Llama3-70B achieve the best performance on Bird dataset. The reason is that for predicted schemas with less redundant information, larger models can better demonstrate their semantic accuracy, thereby achieving more accurate generation of SQL queries.

Table \ref{table3} summarizes the performance of the approaches on the Bird and Spider datasets in terms of schema linking, from which we can observe that PSM-SQL+Llama3-8B achieves the best performance on the MA and RE metrics at the table and column levels at the cost of reducing a bit of score of the IA metric. The reason is that PSM-SQL adopts MSL to learn multi-granularity semantics of schemas and exploits a chain loop strategy to reduce the redundant schemas continuously, benefiting to recognize the related schemas and discard the redundant schemas for schema linking.

\subsection{Ablation Study}
PSM-SQL mainly consists of a chain loop strategy and a MSL module at the column, table, and database levels. In order to verify their effectiveness, we evaluate the variations of PSM-SQL, including PSM-SQL processes one or two rounds of schema linking (denoted as Cycle\_1 and Cycle\_2, respectively), PSM-SQL adopt MSL process schema linking at the column, table, and database level (denoted as Emb\_LLM, Cross\_encoder, and Gen\_LLM, respectively). The performance of the variations is summarized in Table \ref{table4} and Table \ref{table5}, from which we have some observations. 1) Cycle\_2 achieves better performance than Cycle\_1 on SQL generation and schema linking tasks. The reason is that the chain loop strategy can continuously reduce the task difficulty which benefits the model to learn to discard the redundant schemas. 2) MSL achieves better performance than Cross\_encoder, Emb\_LLM, and Gen\_LLM, and performs the best on the IA metric. The reason is that MSL learns the multi-granularity patterns of schemas at the column, table, and database levels, which is comprehensive than a single level. 3) In terms of the performance on SQL generation for inference, Cross\_encoder+Gen\_LLM performs better than MSL. The reason is that Emb\_LLM pays attention on the semantics of columns that is difficult to distinguish, resulting in the retention of a large number of redundant schemas.

\section{Conclusion}\label{sec6}
In this paper, we propose a progressive schema linking with multi-granularity semantics framework with a multi-granularity schema linking module and a chain loop strategy, to link the gold schemas for generating SQL queries from NL questions. More specifically, PSM-SQL employs MSL to capture multi-granularity semantics at column, table, and database levels and uses a chain loop strategy to cyclically reduce the number of schemas, thereby effectively improving the accuracy of schema linking. The experimental results conducted on Spider and Bird demonstrate the effectiveness of the proposed PSM-SQL.

\section{Limitations}\label{sec7}
There are two limitations of our work. Firstly, we did not extensively engineer the prompts to process schema linking and SQL generation, which may not be optimal. Secondly, the chain loop strategy for schema linking inevitably reduces the upper limit of the model while lowering the task difficulty, as it may discard some correct schemas related to SQL generation during cycles, which is a trade-off issue.

\bibliography{custom}

\begin{thebibliography}{24}
\providecommand{\natexlab}[1]{#1}

\bibitem[{Cai et~al.(2021)Cai, Yuan, Xu, and Hao}]{ref3}
Ruichu Cai, Jinjie Yuan, Boyan Xu, and Zhifeng Hao. 2021.
\newblock {SADGA:} structure-aware dual graph aggregation network for
  text-to-sql.
\newblock In \emph{NeurIPS}, pages 7664--7676.

\bibitem[{Cao et~al.(2021)Cao, Chen, Chen, Zhao, Zhu, and Yu}]{ref4}
Ruisheng Cao, Lu~Chen, Zhi Chen, Yanbin Zhao, Su~Zhu, and Kai Yu. 2021.
\newblock {LGESQL:} line graph enhanced text-to-sql model with mixed local and
  non-local relations.
\newblock In \emph{ACL}, pages 2541--2555.

\bibitem[{Deng et~al.(2022)Deng, Chen, and Zhang}]{ref24}
Naihao Deng, Yulong Chen, and Yue Zhang. 2022.
\newblock Recent advances in text-to-sql: {A} survey of what we have and what
  we expect.
\newblock In \emph{COLING}, pages 2166--2187.

\bibitem[{Dong et~al.(2023)Dong, Zhang, Ge, Mao, Gao, lu~Chen, Lin, and
  Lou}]{ref6}
Xuemei Dong, Chao Zhang, Yuhang Ge, Yuren Mao, Yunjun Gao, lu~Chen, Jinshu Lin,
  and Dongfang Lou. 2023.
\newblock \href {https://arxiv.org/abs/2307.07306} {C3: Zero-shot text-to-sql
  with chatgpt}.
\newblock \emph{Preprint}, arXiv:2307.07306.

\bibitem[{Gao et~al.(2024)Gao, Wang, Li, Sun, Qian, Ding, and Zhou}]{ref7}
Dawei Gao, Haibin Wang, Yaliang Li, Xiuyu Sun, Yichen Qian, Bolin Ding, and
  Jingren Zhou. 2024.
\newblock Text-to-sql empowered by large language models: A benchmark
  evaluation.
\newblock In \emph{VLDB}, page 1132–1145.

\bibitem[{Hu et~al.(2022)Hu, Shen, Wallis, Allen{-}Zhu, Li, Wang, Wang, and
  Chen}]{ref22}
Edward~J. Hu, Yelong Shen, Phillip Wallis, Zeyuan Allen{-}Zhu, Yuanzhi Li,
  Shean Wang, Lu~Wang, and Weizhu Chen. 2022.
\newblock Lora: Low-rank adaptation of large language models.
\newblock In \emph{ICLR}, pages 1--13.

\bibitem[{Lee et~al.(2024)Lee, Park, Kim, and Park}]{ref18}
Dongjun Lee, Choongwon Park, Jaehyuk Kim, and Heesoo Park. 2024.
\newblock \href {https://arxiv.org/abs/2405.07467} {Mcs-sql: Leveraging
  multiple prompts and multiple-choice selection for text-to-sql generation}.
\newblock \emph{Preprint}, arXiv:2405.07467.

\bibitem[{Li et~al.(2023{\natexlab{a}})Li, Zhang, Li, and Chen}]{ref2}
Haoyang Li, Jing Zhang, Cuiping Li, and Hong Chen. 2023{\natexlab{a}}.
\newblock {RESDSQL:} decoupling schema linking and skeleton parsing for
  text-to-sql.
\newblock In \emph{AAAI}, pages 13067--13075.

\bibitem[{Li et~al.(2024)Li, Zhang, Liu, Fan, Zhang, Zhu, Wei, Pan, Li, and
  Chen}]{ref12}
Haoyang Li, Jing Zhang, Hanbing Liu, Ju~Fan, Xiaokang Zhang, Jun Zhu, Renjie
  Wei, Hongyan Pan, Cuiping Li, and Hong Chen. 2024.
\newblock Codes: Towards building open-source language models for text-to-sql.
\newblock \emph{ACM on Management of Data}, 2(3):1--28.

\bibitem[{Li et~al.(2023{\natexlab{b}})Li, Hui, Cheng, Qin, Ma, Huo, Huang, Du,
  Si, and Li}]{ref5}
Jinyang Li, Binyuan Hui, Reynold Cheng, Bowen Qin, Chenhao Ma, Nan Huo, Fei
  Huang, Wenyu Du, Luo Si, and Yongbin Li. 2023{\natexlab{b}}.
\newblock Graphix-t5: Mixing pre-trained transformers with graph-aware layers
  for text-to-sql parsing.
\newblock In \emph{AAAI}, pages 13076--13084.

\bibitem[{Li et~al.(2023{\natexlab{c}})Li, Hui, Qu, Yang, Li, Li, Wang, Qin,
  Geng, Huo, Zhou, Ma, Li, Chang, Huang, Cheng, and Li}]{ref14}
Jinyang Li, Binyuan Hui, Ge~Qu, Jiaxi Yang, Binhua Li, Bowen Li, Bailin Wang,
  Bowen Qin, Ruiying Geng, Nan Huo, Xuanhe Zhou, Chenhao Ma, Guoliang Li,
  Kevin~Chen{-}Chuan Chang, Fei Huang, Reynold Cheng, and Yongbin Li.
  2023{\natexlab{c}}.
\newblock Can {LLM} already serve as {A} database interface? {A} big bench for
  large-scale database grounded text-to-sqls.
\newblock In \emph{NeurIPS}, pages 42330--42357.

\bibitem[{Liu et~al.(2019)Liu, Ott, Goyal, Du, Joshi, Chen, Levy, Lewis,
  Zettlemoyer, and Stoyanov}]{ref21}
Yinhan Liu, Myle Ott, Naman Goyal, Jingfei Du, Mandar Joshi, Danqi Chen, Omer
  Levy, Mike Lewis, Luke Zettlemoyer, and Veselin Stoyanov. 2019.
\newblock \href {https://arxiv.org/abs/1907.11692} {Roberta: A robustly
  optimized bert pretraining approach}.
\newblock \emph{Preprint}, arXiv:1907.11692.

\bibitem[{Popescu et~al.(2003)Popescu, Etzioni, and Kautz}]{ref15}
Ana{-}Maria Popescu, Oren Etzioni, and Henry~A. Kautz. 2003.
\newblock Towards a theory of natural language interfaces to databases.
\newblock In \emph{IUI}, pages 149--157.

\bibitem[{Pourreza and Rafiei(2023)}]{ref8}
Mohammadreza Pourreza and Davood Rafiei. 2023.
\newblock {DIN-SQL:} decomposed in-context learning of text-to-sql with
  self-correction.
\newblock In \emph{NeurIPS}, pages 1--10.

\bibitem[{Pourreza and Rafiei(2024)}]{ref11}
Mohammadreza Pourreza and Davood Rafiei. 2024.
\newblock \href {https://arxiv.org/abs/2402.01117} {Dts-sql: Decomposed
  text-to-sql with small large language models}.
\newblock \emph{Preprint}, arXiv:2402.01117.

\bibitem[{Qin et~al.(2022)Qin, Wang, Hui, Li, Wei, Li, Huang, Si, Yang, and
  Li}]{ref23}
Bowen Qin, Lihan Wang, Binyuan Hui, Bowen Li, Xiangpeng Wei, Binhua Li, Fei
  Huang, Luo Si, Min Yang, and Yongbin Li. 2022.
\newblock {SUN:} exploring intrinsic uncertainties in text-to-sql parsers.
\newblock In \emph{COLING}, pages 5298--5308.

\bibitem[{Qu et~al.(2024)Qu, Li, Li, Qin, Huo, Ma, and Cheng}]{ref10}
Ge~Qu, Jinyang Li, Bowen Li, Bowen Qin, Nan Huo, Chenhao Ma, and Reynold Cheng.
  2024.
\newblock \href {https://arxiv.org/abs/2405.15307} {Before generation, align
  it! a novel and effective strategy for mitigating hallucinations in
  text-to-sql generation}.
\newblock \emph{Preprint}, arXiv:2405.15307.

\bibitem[{Talaei et~al.(2024)Talaei, Pourreza, Chang, Mirhoseini, and
  Saberi}]{ref19}
Shayan Talaei, Mohammadreza Pourreza, Yu-Chen Chang, Azalia Mirhoseini, and
  Amin Saberi. 2024.
\newblock \href {https://arxiv.org/abs/2405.16755} {Chess: Contextual
  harnessing for efficient sql synthesis}.
\newblock \emph{Preprint}, arXiv:2405.16755.

\bibitem[{Tang and Mooney(2001)}]{ref16}
Lappoon~R. Tang and Raymond~J. Mooney. 2001.
\newblock Using multiple clause constructors in inductive logic programming for
  semantic parsing.
\newblock In \emph{ECML}, pages 466--477.

\bibitem[{Wang et~al.(2024)Wang, Ren, Yang, Liang, Bai, Chai, Yan, Zhang, Yin,
  Sun, and Li}]{ref9}
Bing Wang, Changyu Ren, Jian Yang, Xinnian Liang, Jiaqi Bai, Linzheng Chai,
  Zhao Yan, Qian-Wen Zhang, Di~Yin, Xing Sun, and Zhoujun Li. 2024.
\newblock \href {https://arxiv.org/abs/2312.11242} {Mac-sql: A multi-agent
  collaborative framework for text-to-sql}.
\newblock \emph{Preprint}, arXiv:2312.11242.

\bibitem[{Wei et~al.(2022)Wei, Wang, Schuurmans, Bosma, Ichter, Xia, Chi, Le,
  and Zhou}]{ref1}
Jason Wei, Xuezhi Wang, Dale Schuurmans, Maarten Bosma, Brian Ichter, Fei Xia,
  Ed~H. Chi, Quoc~V. Le, and Denny Zhou. 2022.
\newblock Chain-of-thought prompting elicits reasoning in large language
  models.
\newblock In \emph{NeurIPS}, pages 24824--24837.

\bibitem[{Xiao et~al.(2024)Xiao, Liu, Zhang, Muennighoff, Lian, and
  Nie}]{ref20}
Shitao Xiao, Zheng Liu, Peitian Zhang, Niklas Muennighoff, Defu Lian, and
  Jian-Yun Nie. 2024.
\newblock \href {https://arxiv.org/abs/2309.07597} {C-pack: Packaged resources
  to advance general chinese embedding}.
\newblock \emph{Preprint}, arXiv:2309.07597.

\bibitem[{Yu et~al.(2018)Yu, Zhang, Yang, Yasunaga, Wang, Li, Ma, Li, Yao,
  Roman, Zhang, and Radev}]{ref13}
Tao Yu, Rui Zhang, Kai Yang, Michihiro Yasunaga, Dongxu Wang, Zifan Li, James
  Ma, Irene Li, Qingning Yao, Shanelle Roman, Zilin Zhang, and Dragomir~R.
  Radev. 2018.
\newblock Spider: {A} large-scale human-labeled dataset for complex and
  cross-domain semantic parsing and text-to-sql task.
\newblock In \emph{EMNLP}, pages 3911--3921.

\bibitem[{Zelle and Mooney(1996)}]{ref17}
John~M. Zelle and Raymond~J. Mooney. 1996.
\newblock Learning to parse database queries using inductive logic programming.
\newblock In \emph{AAAI}, pages 1050--1055.

\end{thebibliography}

\end{document}